\documentclass[twocolumn,eqsecnum,superscriptaddress,showpacs]{revtex4}
\usepackage[dvips]{graphicx}
\begin{document}
\hyphenation{anti-fermion}
\title{  NO AREA LAW IN QCD   }
\vspace{1cm}
\author{T. Asaga}\email{asaga@phys.cst.nihon-u.ac.jp}
\author{T. Fujita}\email{fffujita@phys.cst.nihon-u.ac.jp}

\affiliation{Department of Physics, Faculty of Science and Technology, 
Nihon University, Tokyo, Japan}

\date{\today}%

\begin{abstract}

Wilson's area law in QCD is critically examined. 
It is shown that the expectation value of the Wilson loop integral 
$ \exp( \int iA_\mu dx^\mu ) $ in the strong coupling 
limit vanishes when we employ the conjugate Wilson action 
which has a proper QED action in the continuum limit.  
The finite value of Wilson loop with the Wilson action is due to the result 
of the artifact. The fact that his area law 
is obtained even for QED simply indicates that the area law is unphysical.

\end{abstract}

\maketitle
 
\section{Introduction}

Since Wilson \cite{w1} proposed Wilson loop in evaluating the gauge field 
theories on the lattice, it has been commonly accepted that Wilson's area law 
is the basic conceptual ingredients to understand a confining mechanism 
in QCD. Wilson's formulation of the lattice version of QCD is interesting, 
and in addition it can be used for evaluating the lattice QCD in terms of 
numerical calculations. Indeed, there are many calculations of the lattice 
QCD simulations which confirm the confining mechanism of Wilson's criteria 
\cite{gre}. 

In his paper, Wilson presented an area law which can suggest a confining 
potential of the following type  \cite{w1} 
$$ V(R)  \simeq {\ln g^2\over a^2} R   \eqno{(1)} $$
where $a$ denotes a lattice constant. 
Therefore, it means that the QCD confining potential is a linear rising one. 
Further, this shape of the potential is well reproduced by the lattice QCD 
simulations \cite{creutz}. 

However, one may have somewhat an uneasy feeling on eq.(1). 
That is, eq.(1) does not have a proper dimension since the lattice constant 
(units) cannot play a role of a physical quantity.  In fact, if we take the 
continuum limit of eq.(1), then the right hand side goes to infinity, and 
therefore it becomes unphysical. 
Here, we should explain the continuum limit itself in order to avoid any 
confusions. This is closely related to the way one solves field theory. 
When we wish to solve field theory models, the space and time are of course 
from $-\infty$ to $\infty$. However, we cannot solve the field theory models 
in this space and time, and therefore normally we put the theory into a box 
with its length $L$.  In this case, we can solve field theory models within 
the box and after calculations we should make the length $L$ much larger 
than any other scales of the models, which is called thermodynamic limit. 
If this continuum field theory model can be solved in some way or the other, 
then we can obtain physical observables after the thermodynamic limit is 
taken. 

Sometimes people want to solve the field theory by cutting the space and time 
into a lattice. Wilson's way of solving QCD is just this lattice field theory. 
In this case, we should cut the space and time by the site number $N$, and 
therefore the lattice constant $a$ becomes 
$$ a= {L\over N}    \eqno{(2)} $$
where it is to be noted that the site number $N$ is not a particle 
number of field theory model. 
In this case, it is clear that the lattice constant is not a physical 
quantity since it depends as to how one wishes to cut the space and time. 
In fact, one sees that the potential [eq.(1)] indeed diverges when one wants 
to cut the space as small as possible to simulate the continuum space, 
keeping the box length $L$ finite. Therefore, it is clear that, physically, 
eq.(1) does not make sense. 

Here, it should be important to note that the lattice constant $a$ in solid 
state physics has a completely different meaning from the continuum field 
theory since the constant $a$ of the lattice is fixed and finite. 
In this case, therefore, there is no ambiguity of cutting the space, and 
thus the thermodynamic limit is the only concern in the calculation 
of physical observables. 

In this paper, we show that the finite value of the Wilson loop is an artifact due to 
the Wilson action since it should inevitably pick up unphysical contributions 
to the loop integral. A similar work has been done by Grady \cite{grady} who 
shows that the action different from the Wilson action gives no area law  
in SU(2) lattice gauge theory. There are some works which critically examine lattice 
calculations \cite{cah,pat}, but we do not go into details here. 

\section{ Ambiguities in  averaging procedure }

Now, we should discuss the ambiguities in evaluation of the expectation values. 
The discussion here will be used in the later sections 
to understand the way of evaluating the average of a Wilson loop. 
As examples, we employ the distribution function $ p_{\pm}(A)$ as
$$ p_{\pm}(A)=\exp\left(he^{\pm iA}\right) 
= \sum_{k=0}^\infty {h^k\over{k!}} e^{\pm ikA}   \eqno{(3)}  $$
where $h$ is a constant. This type of the distribution function 
is often used in mathematics in order to pick up some integer 
from certain complicated functions. 

Now, we make an expectation value of $A$ and obtain  
$$ \langle  A \rangle_{\pm} = \sum_{k=0}^\infty {h^k\over{k!}} 
\int_{-{X}}^{X} A e^{\pm ikA}dA / \int_{-{X}}^{X}  p_{\pm}(A) dA =0 
\eqno{(4)}  $$
where $X$ should be eventually set to infinity. 
Eq.(4) shows that there is no finite expectation value of $A$ 
with the distribution function of $p_{\pm}(A)$ in eq.(3). 

However, if we make  expectation values of $e^{ inA}$ where $n$ is a 
positive integer, we obtain
$$ \langle e^{inA} \rangle_{+} =  \sum_{k=0}^\infty {h^k\over{k!}} 
\int_{-{X}}^{X} e^{inA+ikA}dA /
\int_{-{X}}^{X}  p_{+}(A) dA=0  \eqno{(5)}  $$
$$ \langle e^{inA} \rangle_{-} =  \sum_{k=0}^\infty {h^k\over{k!}} 
\int_{-{X}}^{X} e^{inA-ikA}dA /
\int_{-{X}}^{X}  p_{-}(A) dA= {h^n\over{n!}} . \eqno{(6)} $$
The expectation value of $ \langle e^{inA} \rangle_{-} $ 
survives even though the expectation value of $ \langle e^{inA} \rangle_{+} $ 
vanishes. The last result [eq.(6)] shows just the same trick as Wilson used 
in obtaining a finite expectation value of the loop integral, and one notices 
that the procedure is just to pick up the integer value  $n$ in the function 
of $  e^{inA}  $.

\section{ Wilson's  area law }

Now, we come to the area law in Wilson's paper. Most of the lattice 
formulations of the field theory are well written in his paper, and QCD 
on a lattice itself is quite interesting. 
In particular, he introduced Wilson's loop and 
evaluated it in terms of Euclidean path integral formulation. 
Here, we repeat the discussion in his paper using the same notations. 
The expectation value of the Wilson loop is written as  
$$ \langle \  \exp \left( i\sum_P   B_{n\mu} \right) \  \rangle_{+} 
= Z^{-1}_{+} \prod_{m}\prod_\mu \int_{-\pi}^{\pi}dB_{m \mu} $$
$$ \times \exp \left( i\sum_P   B_{n\mu}+{1\over 2g^2}
\sum_{n \mu \nu}e^{ if_{n \mu \nu}} \right)   \eqno{(7)} $$
where the partition function  $Z_{+}$ is defined as
$$ Z_{+}= \prod_{m}\prod_\mu \int_{-\pi}^{\pi}dB_{m \mu}
\exp \left( {1\over 2g^2}\sum_{n \mu \nu}e^{i f_{n \mu \nu}} 
\right) .  \eqno{(8)} $$
$P$ denotes a path on the lattice.
Here, the dimensionless quantity $B_{n \mu} $ is related to the discretized 
vector potential $A_{n \mu}$ as 
$$  B_{n \mu}= ag A_{n \mu}  . \eqno{(9)} $$
A dimensionless form of the field strength $F_{n \mu \nu}$ 
is defined as
$$ f_{n \mu \nu} = a^2 g F_{n \mu \nu} =
B_{n\mu}+B_{n+\hat{\mu},\nu}-B_{n+\hat{\nu},\mu}-B_{n\nu} . 
\eqno{(10)} $$
Wilson imposed the periodic boundary conditions on the vector field 
$A_{n \mu}$,  
$$ A_{n, \mu} =A_{n+N, \mu}    $$
where $N$ is related to the box length $L$ by $L=Na$. 

In the strong coupling limit, eq.(7) can be expressed as
$$  \langle \  \exp \left( i\sum_P   B_{n\mu} \right) \  \rangle _+ 
 = Z_{+}^{-1}\sum_k {1\over k!}\left({1\over 2g^2}\right)^k  $$
$$ \times \prod_{m}\prod_\mu \int_{-\pi}^{\pi}dB_{m \mu} 
 \sum_{\ell_1 \pi_1 \sigma_1} \cdots  
\sum_{\ell_k \pi_k \sigma_k} $$
$$ \times \exp \left( i\sum_P   B_{n\mu}+if_{\ell_1 \pi_1 \sigma_1}+\cdots 
+ if_{\ell_k \pi_k \sigma_k}\right)  . \eqno{(11)} $$
In the same way as eq.(6), eq.(11) has a finite contribution 
only when the exponent in the integrand is zero.  
Therefore, the nonzero terms in the sum 
are those for which satisfy 
$$ \sum_P   B_{n\mu}+f_{\ell_1 \pi_1 \sigma_1}+\cdots 
 +f_{\ell_k \pi_k \sigma_k} =0 . \eqno{(12)} $$

Now, if we specify the path $P$ which contains $K$ plaquette, then we should 
find the number of $f_{\ell_k \pi_k \sigma_k} $ with $k=K$. In this case, 
we see that  eq.(12) is indeed satisfied. Therefore, the $k$ is the number 
of the plaquette that are surrounded by the path $P$.  If this area is 
denoted by $A$, then we find that $k=K={A\over a^2} $.  

Therefore, it is easy to find 
$$  \langle \  \exp \left( i\sum_P   B_{n\mu} \right) \  \rangle _+ 
 \ \sim (g^2)^{-A/a^2}  \eqno{(13)} $$
where $A$ should be described by some 
physical quantity like $A =RT$ with $T$ time distance in Euclidean space. 

From eq.(13), one obtains eq.(1) using the method of transfer matrix which 
relates the Wilson loop to the potential \cite{kog,creut2}. 
However, eq.(13) has a physically improper expression. 
That is, the area $A$ is described in terms of the physical 
quantity while $a^2$ is not a physical quantity since it should be 
eventually put to zero. 
Therefore, it is by now clear that the right hand side cannot survive when 
one takes the continuum limit.

\section{ A  trick  in  Wilson's  calculation }

Why did he obtain such an unphysical result ? 
The answer is simple. The problem is related to the ambiguity in 
the definition of the partition function (8). 
Wilson's area law is obtained only when the Wilson action is employed. 
All other actions including Grady's action \cite{grady} cannot reproduce the area law. 
We will show it below by a simple action with a negative sign 
in the exponential. 

In eq.(8) the gauge field action is expressed as 
$ S_{+}={1\over 2g^2}\sum_{n \mu \nu}e^{i f_{n \mu \nu}}$ 
instead of the conventional form 
$- {1\over 4}a^4\sum_{n \mu \nu}F_{n \mu \nu}^2$. 
For small $a$, this term can be expanded as
$$ S_{+}={1\over 2g^2}\sum_{n \mu \nu}e^{i f_{n \mu \nu}} \approx 
{1\over 2g^2}\sum_{n \mu \nu} (1 + if_{n \mu \nu}-{1\over 2}f_{n \mu \nu}^2 
\cdots ) . \eqno{(14)} $$
The first term is irrelevant and the linear term in $f_{n \mu \nu}$ 
gives no contribution in the calculation. The quadratic term corresponds to 
$- {1\over 4}a^4\sum_{n \mu \nu}F_{n \mu \nu}^2$
which is the conventional expression of the gauge field action. 
Here, one can easily notice that it is also possible to write the action 
for the gauge field with a minus sign in the exponential as 
$S_{-} =  {1\over 2g^2}\sum_{n \mu \nu}e^{- i f_{n \mu \nu}}$. 
Because $S_{-}$ also reduces to 
$S_{-} \sim - {1\over 4}a^4\sum_{n \mu \nu}F_{n \mu \nu}^2$ for small $a$. 
Therefore there is an ambiguity in the definition of $Z$.

The partition function defined with $S_{-}$ is written as 
$$ Z_{-}= \prod_{m}\prod_\mu \int_{-\pi}^{\pi}dB_{m \mu}
\exp \left( {1\over 2g^2}\sum_{n \mu \nu}e^{ - i f_{n \mu \nu}} 
\right) .  \eqno{(15)} $$
In the strong coupling limit, the expectation value of 
the Wilson loop with $ Z_{-}$ becomes 
$$  \langle \  \exp \left( i\sum_P   B_{n\mu} \right) \  \rangle _- 
 = Z_{-}^{-1}\sum_k {1\over k!}\left({1\over 2g^2}\right)^k  $$
$$ \times \prod_{m}\prod_\mu \int_{-\pi}^{\pi}dB_{m \mu} 
 \sum_{\ell_1 \pi_1 \sigma_1} \cdots  
\sum_{\ell_k \pi_k \sigma_k} $$
$$ \times \exp \left( i\sum_P   B_{n\mu}-if_{\ell_1 \pi_1 \sigma_1}
-\cdots -if_{\ell_k \pi_k \sigma_k} \right)  . \eqno{(16)} $$
As in eq.(11), the non-vanishing terms in eq.(16) are those for which satisfy 
$$ \sum_P   B_{n\mu}-f_{\ell_1 \pi_1 \sigma_1}-\cdots 
 -f_{\ell_k \pi_k \sigma_k} =0 . \eqno{(17)} $$
However, for the same path $P$ as in eq.(11), 
there is no term in eq.(16) that satisfies 
the above condition. 
Any set of $-f_{\ell \pi \sigma}$in eq.(16) cannot produce a loop 
in the opposite direction to the path $P$, 
and therefore cannot cancel $ \sum_P B_{n\mu}$.
The situation is the same as eq.(5).
Thus the average of the Wilson loop vanishes to zero 
$$ \langle \exp \left(i \sum_P   B_{n\mu} \right)  \rangle_{- } = 0 .  
\eqno{(18)} $$
Eqs.(13) and (18) show that for a given path $P$ 
$\langle \exp (i \sum_P B_{n\mu} )\rangle_{+}$ and 
$ \langle \exp (i \sum_P B_{n\mu} ) \rangle_{-} $ 
behaves differently. 
This is because it is impossible to satisfy both eqs.(12) 
and (17) simultaneously for a given path $P$. 

Here, we should make a comment on the choice of the action Wilson employed. 
There is some argument that the action should be Hermitian, and therefore 
one should take the action of $(S_{+}+S_{-})/2 $ which is indeed Hermitian. 
However, this modification of the action is a matter of no significance 
since the important part of the action is of course 
$- {1\over 4}a^4\sum_{n \mu \nu}F_{n \mu \nu}^2$ for small $a$. 
Therefore, if the Hermiticity of the action is an important factor in the evaluation 
of physical observables, then it indicates that the action of eq.(14) as well as the 
Hermitian action is physically not a proper one.  

Further, if one calculates the average using 
the conventional action of the gauge field,
$- {1\over 4}a^4\sum_{n \mu \nu}F_{n \mu \nu}^2$, 
then one obtains a different result which has no area law. 
In addition, if one employs an action, for example, 
$ S'={1\over 8g^2}\sum_{n \mu \nu}e^{i 2f_{n \mu \nu}}$, which has a proper 
continuum limit, then one obtains no contribution to the Wilson loop integral.   

From the above discussion, it is 
concluded that the  Wilson loop integral [eq.(13)] is accidentally finite 
as an artifact of the averaging procedure. 
This also indicates that the potential 
$V(R) \simeq {\ln g^2\over a^2}R$ 
obtained from eq.(13) has nothing to do with physics, 
which confirms our point in the first section.

\section{ What is then confinement mechanism }

We have proved that Wilson's area law has nothing to do with 
the confinement in QCD.  The basic mistake of Wilson's calculation is due to 
his action which inevitably picks up unphysical contributions even though 
Wilson's action can be reduced to a proper gauge field action in the continuum 
limit. However, higher order terms of his action contribute nonperturbatively 
to the Wilson loop integral and generate a fictitious area law as a result. 
Therefore, the "confining  potential" which has been believed to be 
confirmed by the numerical calculations of the lattice QCD should be 
reexamined. 

From the dynamical point of view, it is obvious that 
gluon dynamics without fermions cannot give the confinement mechanism 
since gluons produce interactions between quarks which should be 
always Coulomb like potential. Thus, one has to consider quark and 
anti-quark pairs in the intermediate states since they have masses and 
thus can induce a new scale. 
Therefore, physics of the confinement should be obtained only 
after one solved the QCD dynamics with quarks and gluons together. 
In this respect, when one carries out the lattice simulations of QCD, 
the quenched approximation is clearly meaningless. One should take into 
account the fermion degrees of freedom properly. 

Now, what should be a new mechanism of confining  quarks ? 
The confinement itself must be closely related to the gauge invariance 
of physical observables. The color non-singlet objects should be gauge 
dependent, and therefore they cannot be physically observed. 
The quark confinement should be due to the gauge non-invariance of 
the color charge and, because of it, the dynamical confinement should 
occur accordingly. 
In this respect, the fact that Wilson's area law is proved for the U(1) 
gauge theory as well clearly shows that his calculated result 
should not have any relevance to the confinement mechanism in QCD \cite{smit}. 

In this sense, it is most probable that the essence of the confinement may well be 
realized by the MIT bag model \cite{MIT}.

\vspace{0.5cm}

We thank H.A. Weidenm\"uller and M. Yamanaka for useful comments and encouragements. 
We are also grateful to R. Woloshyn for critical comments and for drawing 
our attention to some references.

\end{document}